\documentclass[aps,prl,twocolumn,showpacs,superscriptaddress]{revtex4-1}
\usepackage[usenames,dvipsnames]{color,xcolor}
\usepackage{graphicx,color,epsfig,amssymb,amsmath,lineno,hyperref}
\hypersetup{colorlinks, citecolor=Red, linkcolor=Blue, urlcolor=Red}

\begin{document}
\title{The local load sharing fiber bundle model in higher dimensions}

\author{Santanu Sinha}
\email{Santanu.Sinha@ntnu.no}
\affiliation{Department of Physics, University of Oslo, P. O. Box 1048
  Blindern, N-0316 Oslo, Norway}
\author{Jonas T.\ Kjellstadli}
\email{jonastk@stud.ntnu.no}
\author{Alex Hansen}
\email{Alex.Hansen@ntnu.no}
\affiliation{Department of Physics, Norwegian University of Science and
Technology, N-7491 Trondheim, Norway}
\date{\today}

\begin{abstract}
We consider the local load sharing fiber bundle model in one to
five dimensions. Depending on the breaking threshold
distribution of the fibers, there is a transition where the fracture
process becomes localized. In the localized phase, the model behaves
as the invasion percolation model. The difference between the local
load sharing fiber bundle model and the equal load sharing fiber
bundle model decreases with increasing dimensionality as a power law.
\end{abstract}
\maketitle

The fiber bundle model has come a long way since its introduction in
1926 by Peirce \cite{p26}.  Initially introduced to model the strength
of yarn, the model has slowly gained ground as a fundamental model for
fracture in somewhat the same way that the Ising model has become a
paradigm for magnetic systems. In 1945, the presentation by Daniels
\cite{d45} on the fiber bundle model as a statistical problem led to a
continuous interest for the model in the mechanics community. The
statistical physics community ``discovered" the model in the early
nineties in the aftermath of the surge of interest in fracture and
breakdown phenomena in that community \cite{phc10,hhp15}.

The fiber bundle model introduced by Peirce \cite{p26} is today known
as the {\it equal load sharing\/} (ELS) fiber bundle model. $N$
Hookean springs --- fibers --- of length $x_0$ and spring constant
$\kappa$ are placed between two parallel infinitely stiff clamps. When
the distance between the clamps is $x_0+x$, each fiber carries a load
$\sigma=\kappa x$. Each fiber $i$ has a maximum elongation threshold
$x_i$, upto which it can sustain before failing permanently. The
threshold elongation is drawn from a probability density $p(x_i)$.
The corresponding maximum load that fiber $i$ can sustain is therefore
$\sigma_i=\kappa x_i$.  When a fiber fails, its load is shared equally
among all the surviving fibers since the clamps are infinitely stiff
--- hence the expression ``equal load sharing."

The {\it local load sharing\/} (LLS) fiber bundle model was introduced
by Harlow and Phoenix \cite{hp78,hp91} as a one-dimensional array of
fibers, each having an independent breaking threshold drawn from some
threshold distribution $p(x)$.  They defined the force redistribution
rule as follows: When a fiber fails, the load it carried is
redistributed in equal portions onto its two nearest surviving
neighbors. Hence, if a fiber $i$ is adjacent to $n_{l,i}$ failed
fibers to the left and $n_{r,i}$ failed fibers to the right, it will
carry a load \cite{khh97}
\begin{equation}
\label{eq1}
\sigma_i=\kappa \left[1+\frac{n_{l,i}+n_{r,i}}{2}\right]\ x\;.
\end{equation}

Whereas the ELS fiber bundle model is extreme in the
sense that it redistributes the force carried by the failed fibers
equally among all surviving fibers wherever they are placed, the local
load sharing fiber bundle model is extreme in the opposite sense: only
the nearest survivors, pick up the force carried by the failed fiber.
There are many models that are intermediate between the two extreme
models. For example, the $\gamma$ model of Hidalgo et al.\
\cite{hmkh02} distributes the force carried by the failed fiber
according to a power law in the distance from the failed fiber. The
soft clamp model \cite{bhs02,sgh12,gsh13,gsh14} replaces one of the
infinitely stiff clamps in the ELS model by a clamp with
finite elastic constant. Hence, the redistribution of the load of a
failed fiber is governed by the elastic response of the soft clamp.

We emphasize the following subtle point in the implementation of the
LLS model \cite{hhp15}. If the redistribution of forces
after the failure of a fiber proceeds by dividing the force it carried
in two and adding each half to the two nearest surviving fibers to the
left and right --- i.e., according to the recipe of Harlow and Phoenix
\cite{hp78} --- the force distribution will not follow
Eq. (\ref{eq1}). Rather, it will become dependent on the order in
which the fibers have failed. Hence, it will not be possible to
determine the force distribution among the fibers only from the
knowledge of present failed fibers in the system. This history
dependency in the force distribution is unphysical. To give an
example, if two adjacent fibers have failed and the two nearest
surviving fibers each has one survivor, the first procedure will
produce the following loads on the fibers: $(7/4,0,0,9/4)$ or
$(9/4,0,0,7/4)$ depending on the order in which the two middle fibers
were failed. According to Eq. (\ref{eq1}), the load distribution
should be $(2,0,0,2)$ independent of the order in which the fibers
failed.

When implementing the LLS model in two or more
dimensions, the algorithm by which the forces are redistributed
becomes even more crucial. We insist that the model should be physical
where the force distribution among the surviving fibers can be
determined by only knowing which fibers are already failed and it should
not depend on the order in which they have failed. This leads to the
concept of clusters of failed fibers, where the term ``cluster" is
used in the same sense as in the site percolation problem \cite{sa94}:
failed fibers that are nearest neighbors to each other form a cluster. The
total load carried by all the failed fibers in a given cluster will
then be shared equally by the surviving fibers that form the 
perimeter to that cluster. If a surviving fiber is
adjacent to two different clusters of failed fibers, the total load it
carries is the sum of the loads contributed from both the clusters.

\begin{figure}
\centerline{
\psfig{file=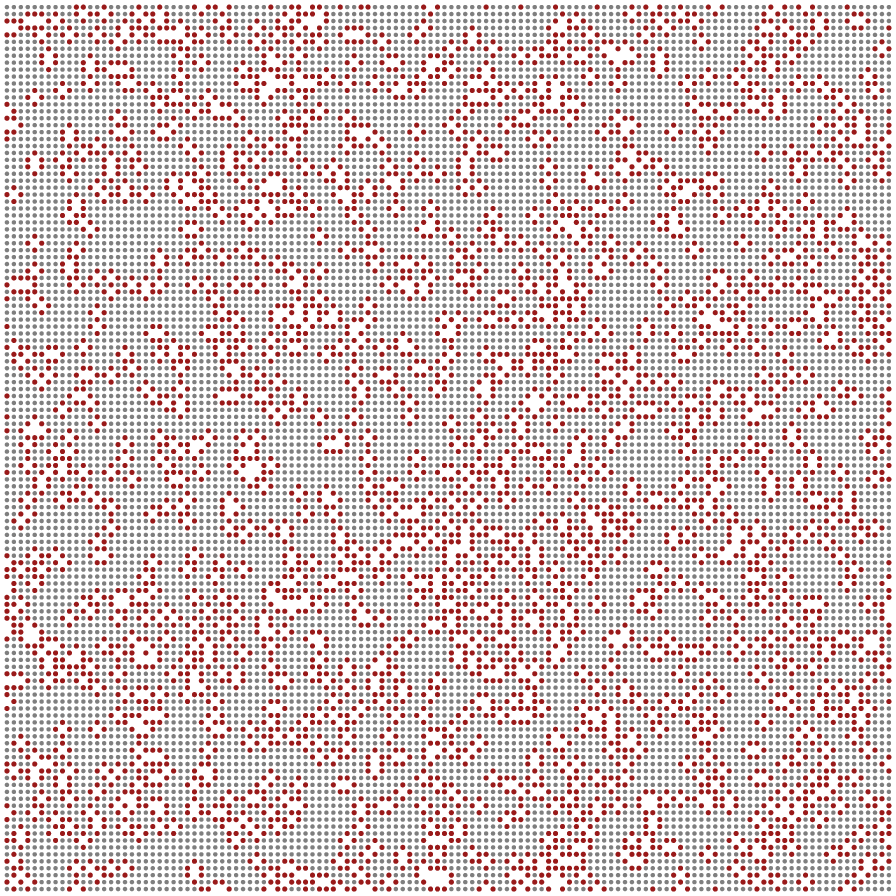,width=0.15\textwidth}\hfill
\psfig{file=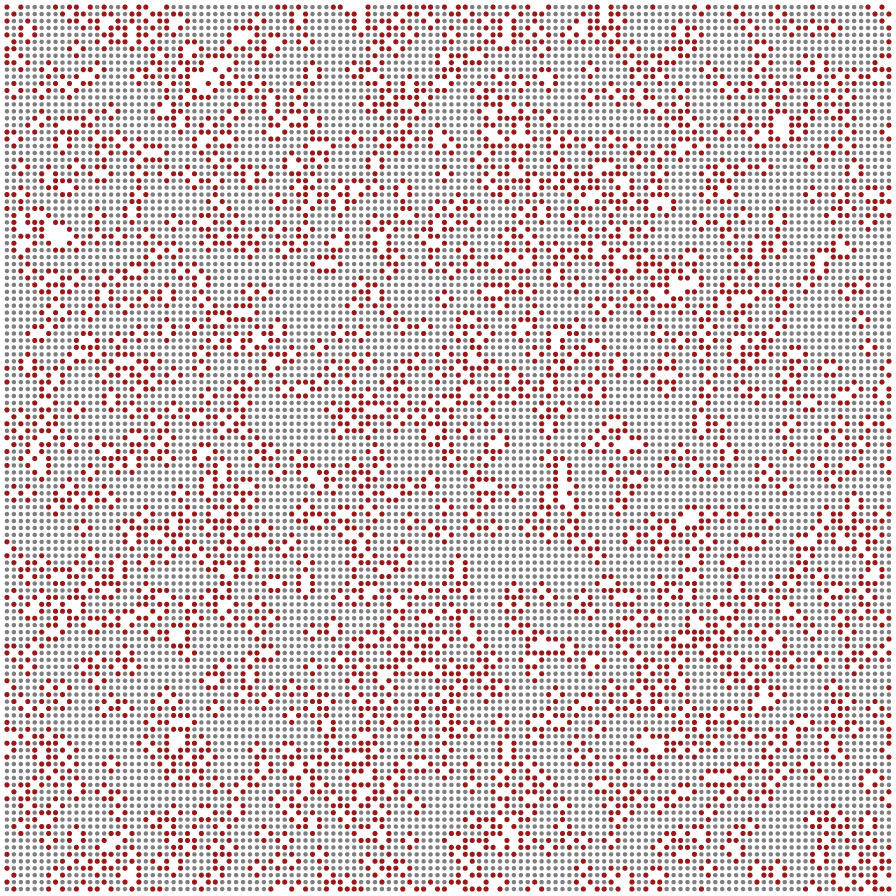,width=0.15\textwidth}\hfill
\psfig{file=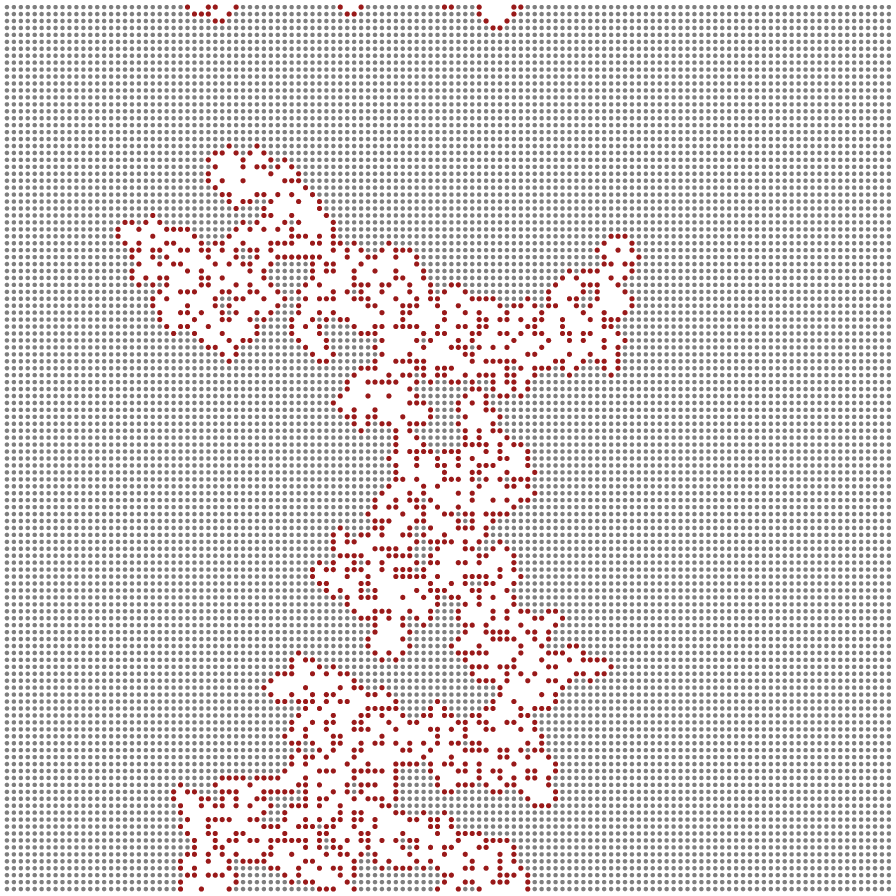,width=0.15\textwidth}}
\medskip
\centerline{
\psfig{file=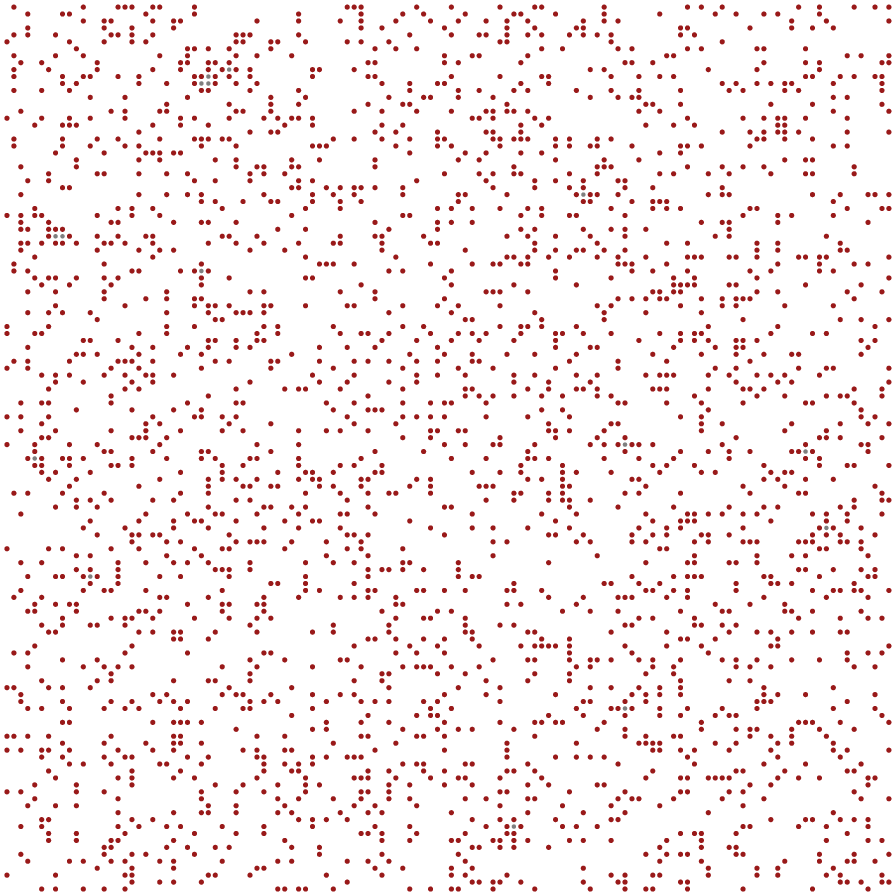,width=0.15\textwidth}\hfill
\psfig{file=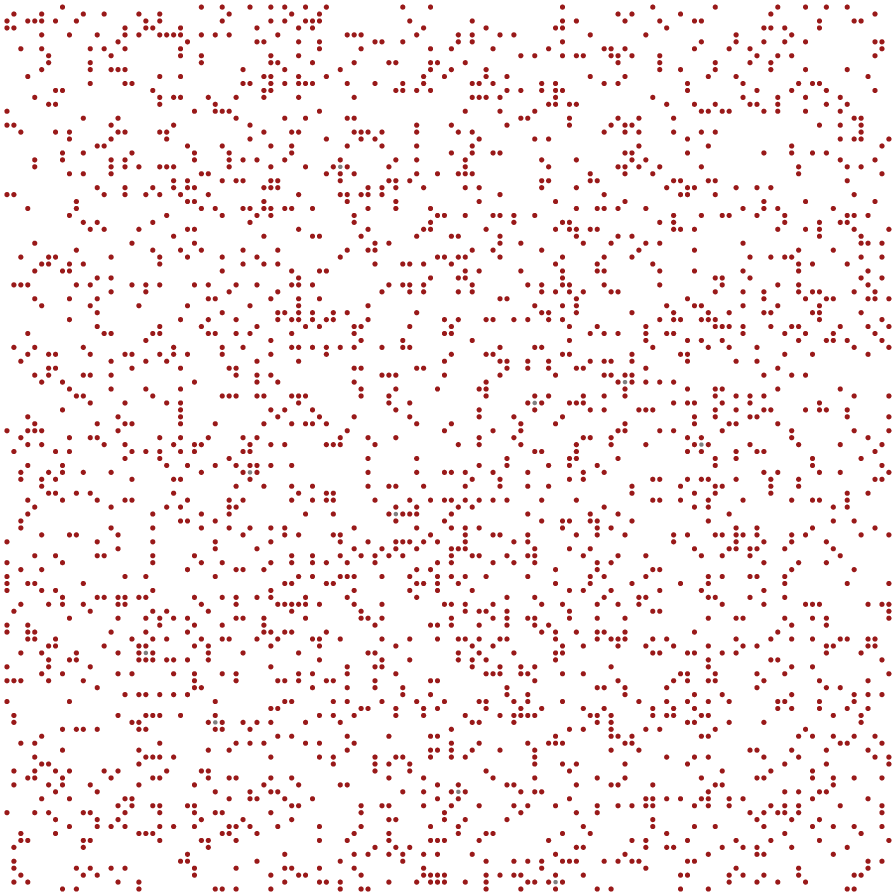,width=0.15\textwidth}\hfill
\psfig{file=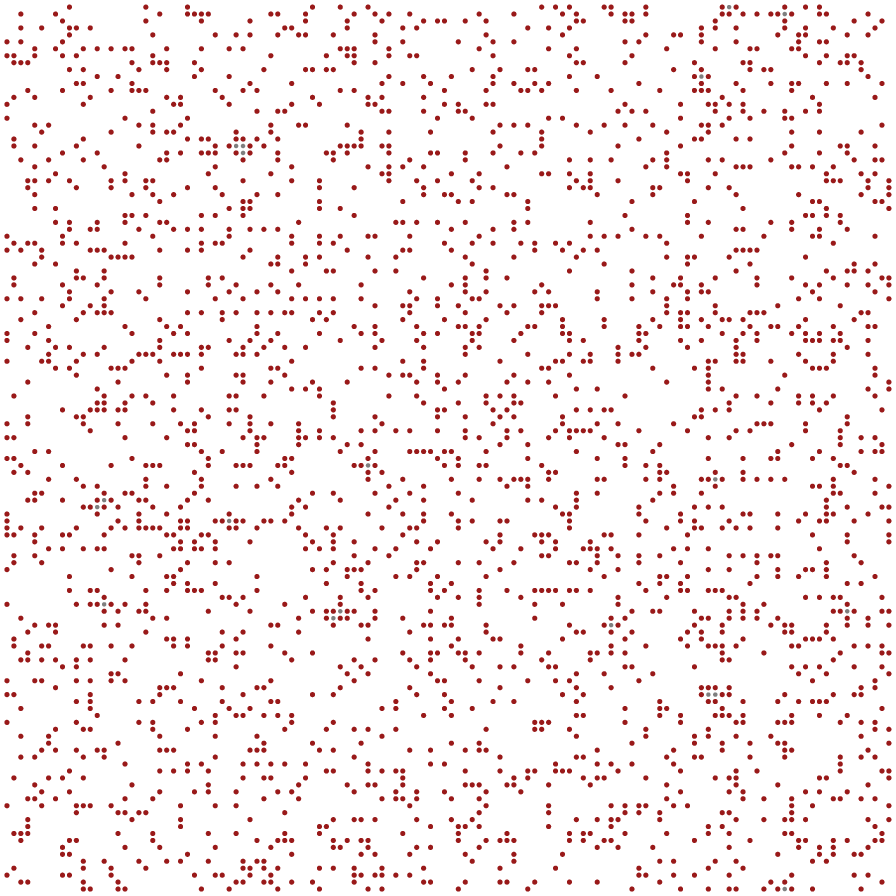,width=0.15\textwidth}}
\centerline{\footnotesize \hfill$p(x)=1$ \hfill \hfill $p(x)=\exp(-x)$
\hfill $p(x)=\exp(-x+1)$}
\caption{\label{fig1} (Color online) Snapshots of the two-dimensional
LLS model after $1792$ failed fibers (top row) and after $13824$
failed fibers (bottom row). The first column shows a uniform
threshold distribution on the unit interval, the middle column is an
exponential threshold distribution $p(x)=\exp(-x)$ where
$x\in[0,\infty)$. The third column shows the threshold distribution
$p(x)=\exp(-x+1)$ where $x\in [1,\infty)$. The system size is
$128\times 128$. The red fibers are survivors adjacent to clusters
of failed fibers, the gray fibers are survivors that are not
adjacent to failed fibers and white fibers have failed.}
\end{figure}

This generalization of the one-dimensional LLS model to higher
dimensions is the simplest one that ensures history independence in
the force distribution. A more elaborate generalization may be found
in Patinet et al.\ \cite{pvhr14}. Here, one of the clamps is exchanged
for a stretchable membrane that has no bending resistance. The elastic
response of this model is equivalent to the LLS model in one
dimension. However, it differs from the one we propose here in two
dimensions.

We show in Fig.\ \ref{fig1} the two stages of the two-dimensional local
load sharing model: after $1792$ failed fibers (top row) and after
$13824$ failed fibers (bottom row). The total number of fibers was
$N=128^2$. The fibers, placed at the nodes of a square lattice, are
seen from above. The failed fibers are shown as white, the intact
fibers that belong to the external and internal perimeters of the 
clusters of failed fibers are shown as red. The intact fibers that 
do not belong to the perimeters are shown as gray. There are periodic 
boundary conditions in all directions. In the first column of the figure, 
the threshold distribution $p(x)$ was uniform on the unit interval.
Hence, the cumulative probability was $P(x)=x$ where $x\in [0,1]$. In
the next two columns, the cumulative threshold probability was
$P(x)=1-\exp(x_<-x)$ where $x\in [x_<,\infty)$. In the middle column,
$x_< = 0$ and in the third column $x_<=1$. In the top row, it is hard
to distinguish the difference between the first two panels of the
figure. However, the third panel in the top row is very different.
In this case, the breakdown process is {\it localized\/} from
the very beginning. That is, a single cluster of failed fibers
forms and keeps growing. On the other hand, three panels in the
bottom row are all very similar.

\begin{figure}
\centerline{\psfig{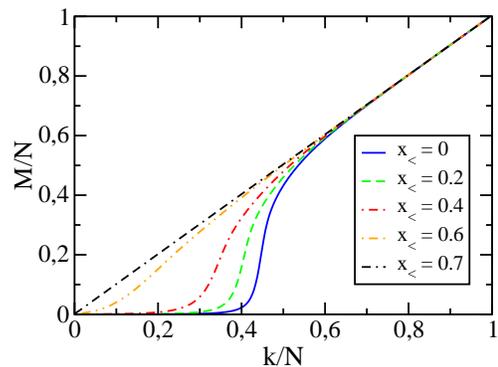}}
\caption{\label{fig2} (Color online) The size of the largest hole in
the two-dimensional LLS model as a function of the
relative number of failed fibers $k/N$, where $N=256^2$. The
threshold distribution was $p(x)=\exp(-x+x_<)$, where
$x\in[x_<,\infty)$. Each data set is based on $5000$ samples.}
\end{figure}

When the breakdown process is localized so that only one cluster of
failed fibers forms, the model is equivalent to the {\it invasion
percolation model\/} \cite{ww83}. In the invasion percolation model,
each site is given a random number. An initial site is invaded. The
perimeter of this one-site cluster form the growth sites and the
growth site with the smallest random number associated with it is
invaded. This is repeated, letting the perimeter of the cluster of
invaded sites to be the growth sites. In the LLS model,
the perimeter of the single cluster of failed fibers will carry the
extra force that makes these and only these fibers liable for failure
when the threshold is narrow enough to imply localization. It will be
the fiber in the perimeter that has the smallest failure threshold
that will fail next. Hence, it behaves precisely as the invasion
percolation model.

The onset of localization is illustrated in Fig. \ref{fig2}. Here we
show the size of the largest cluster of failed fibers, $M$ as a
function of the number of failed fibers $k$ for different values of
the lower cutoff $x_<$ in the threshold distribution $p(x) =
\exp(x_<-x)$ where $x\in[x_<,\infty)$. When $x_<=0.7$, the size of the
largest cluster grows linearly with $k$ from the very beginning,
implying localization. On the other hand, when $x_<=0$, the size of
the largest cluster remains very small for a long period, and then
grows rapidly afterwards. This is caused by merging of smaller
clusters. However, the value at which the derivative of the curve
is the largest is considerably smaller than the value $k/N \approx
0.592746$, the site percolation threshold \cite{sa94},
emphasizing that the failure process is {\it not\/} a random
percolation process.

\begin{figure}
\centerline{\psfig{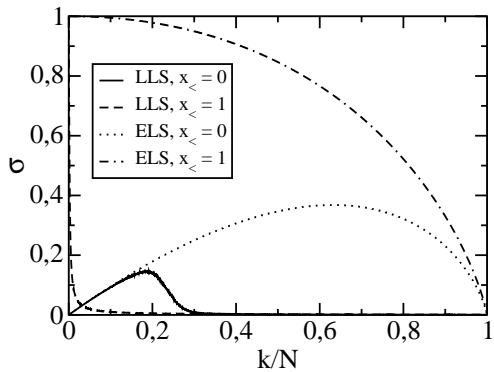}}
\caption{\label{fig3} Force per fiber $\sigma$ as a function of the
relative number of failed fibers $k/N$ in the one-dimensional local
load sharing model together with the ELS model result
(Eq. \ref{eq5}). The thresholds were distributed according to $p(x)
= \exp(-x+x_<)$ where $x\in[x_<,\infty)$ was used with $x_<=0$ and
$1$ respectively. Here $N=4000$. Each data series is based on
$2000$ samples.}
\end{figure}

We now consider the breaking characteristics of the LLS model in
comparison to the ELS model. When $k$ fibers have failed, the force
$F$ carried by the surviving fibers in the ELS model is
\begin{equation}
\label{eq2}
F=N\sigma=(N-k)\ \kappa\ x\;,
\end{equation}
where we have defined the force per fiber $\sigma=F/N$. In the local
load sharing model, we have
\begin{equation}
\label{eq3}
F=N\sigma= N\ \kappa\ x\;,
\end{equation}
since the surviving perimeter fibers precisely absorb the load carried
by the failed fibers.

We order the failure thresholds of the $N$ fibers in an ascending
sequence, $x_{(1)}<x_{(2)}<\cdots < x_{(m)}<\cdots
<x_{(N)}$. According to order statistics \cite{g04}, the average (over
samples) of the $m$th member of this sequence is given by
\begin{equation}
\label{eq4}
P\left(\langle x_{(m)}\rangle\right) = \frac{m}{N}\;
\end{equation}
for large $N$.
We combine this equation with Eq.\ (\ref{eq2}) for the 
ELS model assuming that $P(x)=1-\exp(-x+x_<)$ for
$x\in[x_<,\infty)$ to find
\begin{equation}
\label{eq5}
\sigma=\left[1-\frac{k}{N}\right]\ 
\left[x_<-\ln\left(1-\frac{k}{N}\right)\right]\;.
\end{equation}
For a uniform threshold distribution in $[x_<,1]$, the cumulative
probability is $P(x)=(x-x_<)/(1-x_<)$ we find
\begin{equation}
\label{eq5.1}
\sigma=\left[1-\frac{k}{N}\right]\ \left[x_<+(1-x_<)\frac{k}{N}\right]\;.
\end{equation}
We show the ELS behavior for the exponential threshold distribution
(Eq. \ref{eq5}) in Fig.\ \ref{fig3} together with the corresponding
curves ($x_<=0$ and $x_<=1$) for the LLS model in one dimension. There
is a large difference between ELS and LLS models.

\begin{figure}
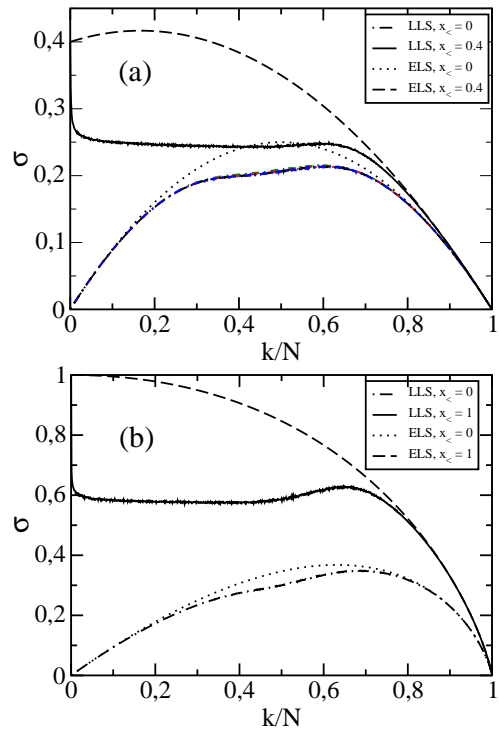

\psfig{file=fig4a.eps,width=0.36\textwidth}
\psfig{file=fig4b.eps,width=0.36\textwidth}
\caption{\label{fig4} (Color online) Force per fiber $\sigma$ as a
function of the relative number of failed fibers $k/N$ in the
two-dimensional LLS model compared with the ELS
model result. The threshold distribution in (a) was
uniform in the interval $x\in[x_<,1]$ with $x_<=0$ and $0.4$. In
(b), the distribution $p(x) = \exp(x_<-x)$ where $x\in[x_<,\infty)$
with $x_<=0$ and $1$ respectively. In (a) for LLS, results are also
shown for different system sizes, $N=32^2$ (green), $64^2$ (red),
$128^2$(black), and $256^2$ (blue) for the uniform distribution with 
$x_<=0$. For the other plots, $N=256^2$. Each data series is based on
$5000$ samples.}
\end{figure}

This picture changes in two dimensions. In Fig.\ \ref{fig4}, we show
the results for the two-dimensional LLS model for uniform threshold
distribution with cumulative threshold probabilities
$P(x)=(x-x_<)/(1-x_<)$, where $x\in[x_<,1]$ with $x_<=0$ and $0.4$
in (a). In Fig.\ \ref{fig4}(b), $P(x)=1-\exp(x_<-x)$
where $x\in[x_<,\infty)$ with $x_<=0$ and $1$. When comparing this
figure to the corresponding one for one dimension (Fig. \ref{fig3}),
we see that the LLS model now is much closer to the ELS model 
than in one dimension.

It should be pointed out that $\sigma$ vs.\ $k/N$ for the exponential
threshold distribution with $x_<=1$ has a curious upwards bend before
its maximum value, see Fig.\ \ref{fig4}. A small upwards bend can also
be seen for the uniform threshold distribution with $x_<=0.4$. This
means that the model is stable in this region in the sense that if
$\sigma$ is used as the control parameter, fiber failures will only
occur if $\sigma$ is increased. This is not true in the 
ELS model.  Hence, the LLS model is in fact {\it
more stable\/} than the ELS model in this region. 

\begin{figure}
\centerline{\psfig{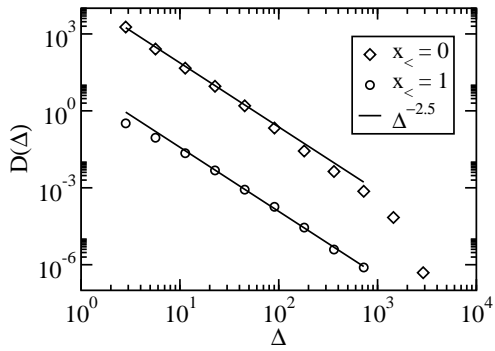}}
\caption{\label{fig5} The burst distribution in the two-dimensional
LLS model. The threshold distribution was $p(x)=\exp(x_<-x)$ where
$x\in[x_<,\infty)$. The data sets are based on $5000$ samples of
size $N=256^2$.}
\end{figure}

The similarity between the ELS and LLS models is also evident 
in other quantities that characterize the two models. In Fig.\ 
\ref{fig5} we show the burst distribution
for the LLS model in two dimensions for the cumulative
threshold probability $P(x)=1-\exp(x_<-x)$ where $x\in[x_<,\infty)$
with $x_<=0$ and $1$. The burst distribution is the histogram of the
number of simultaneously failing fibers $\Delta$ when the force
$\sigma$ is the control parameter. Hemmer and Hansen showed in 1992
that the burst distribution in the ELS model is given by
\begin{equation}
\label{eq6}
\omega(\Delta) \sim \Delta^{-5/2}\;,
\end{equation}
for a very wide class of threshold distributions to which $p(x)=\exp(x_<-x)$ 
belongs \cite{hh92}.

Later, Hansen and Hemmer investigated the burst distribution in the
one-dimensional LLS model finding a burst exponent $\approx 4.5$
rather than $5/2$ \cite{hh94}. Kloster {\it et. al.}  \cite{khh97}
showed analytically that the burst distribution falls off faster than
a power law in the LLS model when the threshold distribution is
uniform on the unit interval. Fig.\ \ref{fig5} shows that the burst
distribution in the two-dimensional LLS model is consistent with
Eq. (\ref{eq6}) for both $x_<=0$ and $x_<=1$.

\begin{figure}
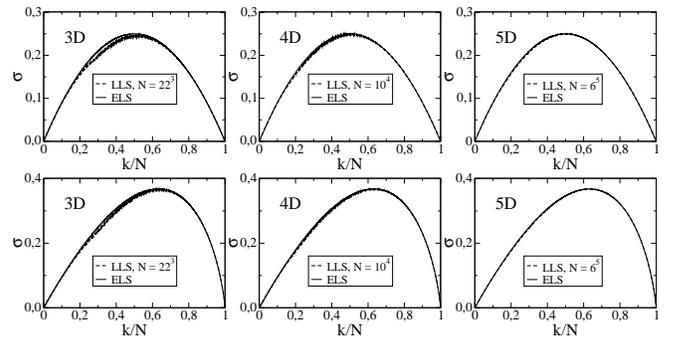

\centerline{
\psfig{file=fig6a.eps,width=0.16\textwidth}\hfill
\psfig{file=fig6b.eps,width=0.16\textwidth}\hfill
\psfig{file=fig6c.eps,width=0.16\textwidth}}
\centerline{
\psfig{file=fig6d.eps,width=0.16\textwidth}\hfill
\psfig{file=fig6e.eps,width=0.16\textwidth}\hfill
\psfig{file=fig6f.eps,width=0.16\textwidth}}
\caption{\label{fig6} Force per fiber $\sigma$ as a function of the
relative number of failed fibers $k/N$ in the three-dimensional
(3D), four-dimensional (4D) and five-dimensional (5D) LLS model. The
top row corresponds to the threshold distributions $P(x)=x$ with
$x\in[0,1]$ and bottom row corresponds to $P(x)=1-\exp(x_<-x)$ with
$x\in[x_<,\infty)$. The system sizes are indicated in the
figures. The number of samples over which the data are averaged
are $80000$, $40000$ and $40000$ for three, four and five
dimensions respectively.}
\end{figure}

\begin{figure}
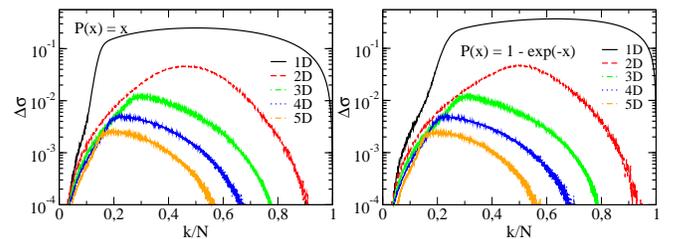

\centerline{
\psfig{file=fig7a.eps,width=0.24\textwidth}\hfill
\psfig{file=fig7b.eps,width=0.24\textwidth}}
\caption{\label{fig7} (Color online) Difference of the force per fiber
$\sigma$ in the LLS model from that in 
ELS model ($\Delta\sigma$) for two threshold distributions
$P(x)=x$ with $x\in[0,1]$ and $P(x)=1-\exp(x_<-x)$ with
$x\in[x_<,\infty)$ in one, two, three, four and five dimensions. A
rapid decrease in $\Delta\sigma$ can be observed with increasing
dimensionality.}
\end{figure}

In Fig.\ \ref{fig6}, we show the $\sigma$ vs.\ $k/N$ curves for the
{\it three-dimensional\/}, {\it four-dimensional\/} and {\it
five-dimensional\/} LLS fiber bundle model for the
cumulative threshold probability $P(x)=x$ with $x\in[0,1]$ (top row)
and $P(x)=1-\exp(x_<-x)$ with $x\in[x_<,\infty)$ (bottom row). We
compare the curves with the ELS model results given in
Eqs.\ (\ref{eq5}) and (\ref{eq5.1}). Interestingly, the curves for
the local and the ELS models are approaching each other
more and more as the dimensionality is increased. The difference in
$\sigma$ for LLS and ELS for different system dimensions is 
measured and plotted in Fig.\ \ref{fig7} for the two threshold
distributions.  

\begin{figure}
\centerline{\psfig{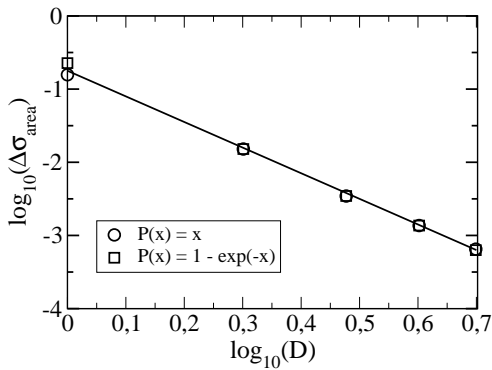}}
\caption{\label{fig8} The area under the $\Delta\sigma$ curves
in Fig.\ \ref{fig7} as a function of dimensionality $D$.}
\end{figure}

It can be noticed that the maxima of the $\Delta\sigma$ curves shifts
towards smaller $k/N$ with changing dimensionality. Therefore, in
order to quantify the difference between LLS and ELS models, we
measure the total area ($\Delta\sigma_\text{area}$) under the
$\Delta\sigma$ curves. In Fig.\ \ref{fig8}, we plot
$\Delta\sigma_\text{area}$ as a function of the dimensionality $D$ of
the system. Interestingly, a power-law dependency 
\begin{equation}
\label{eq7}
\Delta\sigma_{\rm area} \sim D^{-\mu}\;,
\end{equation}
with $\mu=3.5\pm0.1$ is observed.

In the cases where the threshold cutoff is $x_< >0$, there is
a non-negligible $N$-dependency in the $\sigma$ vs.\ $k/N$
curves and the effective exponent $\mu$ needs further finite size
scaling analysis to be determined.

From Eq.\ (\ref{eq7}) we conclude that there is no finite upper 
critical dimension for which the LLS and ELS models become equal. 
However, the difference falls off rapidly with increasing $D$.

Finally, we like to highlight about the breaking process which makes
the LLS and the ELS models similar at the earlier and
later of the breakdown process when there is no localization. The
right column in Fig.\ \ref{fig1} shows the two-dimensional 
LLS model after $13824$ out of $128^2=16384$ fibers in total have
failed, $k/N\approx 0.84$. The clusters of failed fibers have merged
and essentially all the remaining fibers have become part of the
perimeter of a single percolating cluster of failed fibers. Hence, all
the remaining fibers experience the same force as they all are
adjacent to the same cluster --- and hence, they all share the same
force as in the ELS fiber bundle model.

Early in the breakdown process, when there is no localization, fibers
will fail not due to being under stress because they are on the
perimeter of clusters of already failed fibers fibers, but because
they have small thresholds. Hence, early in the breakdown process, we
expect the LLS and the ELS models to be quite similar. This is true in
all dimensions, except when there is localization, see
Figs.\ \ref{fig3}, \ref{fig4} and \ref{fig6}.

The LLS model is extreme in that it is the perimeter fibers that
absorb the forces from the failed fibers. We have mentioned models
that are in between the ELS and the LLS models. When the LLS and ELS
models are rapidly approaching each other with increasing $D$, so will
the in-between models also; they will rapidly approach the ELS model
with increasing $D$. This argument also apply to models that normally
are not classified as fiber bundle models, such as the fuse model
where Zapperi {\it et. al.} \cite{zns05} has reported a burst
distribution exponent in three dimensions equal to $2.55$, close to
the ELS value $5/2$. Hence, already in three dimensions, the ELS model
is not far from the much more complex models of fracture.

We thank Per Christian Hemmer and Srutarshi Pradhan for numerous
discussions on this subject. S.S.\ thanks the Norwegian Research
Council for support through grant 216699.




\begin{thebibliography}{21}

\bibitem{p26} F.\ T.\ Peirce,  J.\ Text.\ Ind.\ {\bf 17}, 355 (1926).

\bibitem{d45}H.\ E.\ Daniels,  Proc.\ Roy.\ Soc.\ London {\bf A183},
405 (1945).

\bibitem{phc10} S.\ Pradhan, A.\ Hansen and B.\ K. Chakrabarti,
Rev.\ Mod.\ Phys.\ {\bf 82}, 499 (2010).

\bibitem{hhp15} A.\ Hansen, P.\ C.\ Hemmer and S.\ Pradhan, {\it The fiber
bundle: modeling failure in materials\/} (J.\ Wiley, Chichester, 2015).

\bibitem{hp78} D.\ G.\ Harlow and S.\ L.\ Phoenix, J.\ Composite Mater.\
{\bf 12}, 195 (1978).

\bibitem{hp91} D.\ G.\ Harlow and S.\ L.\ Phoenix, J.\ Mech.\ Phys.\
Solids {\bf 39}, 173 (1991).

\bibitem{khh97} M.\ Kloster, A.\ Hansen and P.\ C.\ Hemmer, Phys.\
Rev.\ E {\bf 56}, 2615 (1997).

\bibitem{hmkh02} R.\ C.\ Hidalgo, Y.\ Moreno, F.\ Kun and H.\ J.\ Herrmann,
Phys.\ Rev.\ E \textbf{65}, 046148 (2002).

\bibitem{bhs02} G.\ G.\ Batrouni, A.\ Hansen and J.\ Schmittbuhl,
Phys.\ Rev.\ E {\bf 65}, 036126 (2002).

\bibitem{sgh12} A.\ Stormo, K.\ S.\ Gjerden and A.\ Hansen,
Phys.\ Rev.\ E {\bf 86}, R025101 (2012).

\bibitem{gsh13} K.\ S.\ Gjerden, A.\ Stormo and A.\ Hansen,
Phys.\ Rev.\ Lett.\ {\bf 111}, 135502 (2013).

\bibitem{gsh14}  K.\ S.\ Gjerden, A.\ Stormo and A.\ Hansen,
Front.\ Phys.\ {\bf 2}, 66 (2014).

\bibitem{sa94} D.\ Stauffer and A.\ Aharony, {\it Introduction to
percolation theory\/} (Taylor and Francis, London, 1994).

\bibitem{pvhr14} S.\ Patinet, D.\ Vandembroucq, A.\ Hansen and S.\ Roux,
Europ.\ J.\ Phys.\ Spec.\ Top.\ {\bf 223}, 2339 (2014).

\bibitem{ww83} D.\ Wilkinson and J.\ F.\ Willemsen, J.\ Phys.\ A
{\bf 16}, 3365 (1983).

\bibitem{g04} E.\ J.\ Gumbel, {\it Statistics of extremes\/} (Dover
Publ., Mineola New York, 2004).

\bibitem{hh92} P.\ C.\ Hemmer and A.\ Hansen, ASME J.\ Appl.\ Mech.\
{\bf 59}, 909 (1992).

\bibitem{hh94} A.\ Hansen and P.\ C.\ Hemmer, Phys.\ Lett.\ A {\bf 184},
394 (1994).

\bibitem{zns05} S.\ Zapperi, P.\ K.\ V.\ V.\ Nukala and S.\ \v{S}imunovi{\'c},
Physica A {\bf 357}, 129 (2005).

\end{thebibliography}
\end{document}